\documentclass[prl,superscriptaddress,twocolumn]{revtex4}
\usepackage[dvips]{graphicx}
\usepackage{latexsym}
\begin{document}

\newcommand{\fig}[2]{\includegraphics[width=#1]{#2}}

\title{Chaos and residual correlations in pinned disordered systems}
\author{Pierre Le Doussal}
\affiliation{CNRS-Laboratoire de Physique Th\'eorique de l'Ecole Normale Sup\'erieure,
24 Rue Lhomond 75231 Paris,
France}

\date{May 27 2005}

\begin{abstract}
We study, using functional renormalization (FRG), two copies of an
elastic system pinned by mutually correlated random potentials.
Short scale decorrelation depend on a non trivial boundary layer regime
with (possibly multiple) chaos exponents. Large scale mutual
displacement correlation behave as $|x-x'|^{2 \zeta - \mu}$, the
decorrelation exponent $\mu$ proportional to the difference
between Flory (or mean field) and exact roughness exponent
$\zeta$. For short range disorder $\mu >0$ but small, e.g. for
random bond interfaces $\mu = 5 \zeta - \epsilon$, $\epsilon=4-d$,
and $\mu = \epsilon (\frac{(2 \pi)^2}{36} - 1)$ for the one
component Bragg glass. Random field (i.e long range) disorder
exhibits finite residual correlations (no chaos $\mu = 0$)
described by new FRG fixed points. Temperature and dynamic chaos (depinning)
are discussed. 

\end{abstract}

\maketitle

Low energy configurations in
glass phases induced by quenched disorder are non
trivial and their statistics are far from fully characterized.
It was proposed \cite{mckay,braymoore,droplets}
that they exhibit ''static chaos'' i.e. extreme
sensitivity to a small change of external parameters, denoted
$\delta$ (e.g. disorder, temperature, interactions). Suitably
defined correlations, or overlaps, between two (sub)systems are
argued to decay to zero beyond the overlap length
$L_\delta$, which diverges at small $\delta$ as $L_\delta \sim
\delta^{-1/\alpha}$, $\alpha$ the chaos exponent. A
phenomenological droplet description 
predicts \cite{braymoore} $\alpha =
\frac{d_s}{2} - \theta$ where $-\theta$ is the thermal eigenvalue
and $d_s$ the droplet (fractal) dimension, which
should apply to spin glasses as well as to disordered elastic
systems (there $d_s=d$) \cite{droplets}.
Numerical simulations \cite{rieger} have qualitatively at least confirmed this
picture for systems of low dimensions $d$. In higher $d$ 
and in mean field models,
the presence of (temperature) chaos is far more elusive and 
it is still highly debated whether $L_\delta$ is infinite,
or extremely large \cite{nifle}. 
Chaos is a central issue for the
theory of glasses and can be explored in many 
experimental systems either
by varying external parameters or constructing
actual correlated copies 
\cite{chaosexperiments1,chaosexperiments2,chaosexperiments3}.
One wants to develop analytical tools beyond mean field
for this problem, and also investigate
residual correlations at large scale.

Here we focus on the simplest class,
elastic manifolds in random potentials.
Upon an applied force $f$ they exhibit a depinning
transition, also with non trivial rough configurations \cite{depin}. They have
numerous experimental applications for pinning of
e.g. domain walls in magnets \cite{creepexp}, electronic crystals,
density waves \cite{cdw}, vortices in superconductors
\cite{vortices,braggglass}, and wetting \cite{rolley}.
They exhibit energy dominated glass fixed points where
temperature is irrelevant, $T_L \sim L^{- \theta}$, 
$\theta = d - 2 + 2 \zeta$ the free energy
fluctuation exponent, $\zeta$ the roughness exponent of the
pinned configurations $u \sim L^\zeta$. Chaos in pinned manifolds was
studied mostly via scaling arguments \cite{shapir}.
The directed polymer ($d=1$) was studied
numerically and via analytical arguments for $N=1$
indicating $\alpha=1/6$ in agreement with droplets 
\cite{Zhang}, and recently, on 
hierarchical lattices \cite{silveira}.
In $d=2$ chaos was demonstrated for periodic
systems near the glass transition $T_g$ using the Cardy
Ostlund RG, and linked to universal susceptibility fluctuations
as in mesoscopic disordered metals \cite{hwafisher}. 

Pinned manifolds can be studied using functional renormalization
(FRG) within an expansion in $\epsilon=4-d$, to one loop
\cite{fisher,frgdep} and recently to higher orders
\cite{twoloopdep}. In the large $N$ limit
\cite{largeN} it matches results from
mean field theory \cite{mezpar}.
The main aim of this paper is to
investigate static chaos using the FRG. It also opens the way to
investigate ''dynamic chaos'' in driven situations, 
e.g. at depinning $f=f_c$, or in moving phases \cite{mg}, where no 
other analytical tool seems available.
Results also provide a test for the FRG,
as a complete understanding of the field theory (FT)
at low temperature remains a challenge. In particular 
we unveil a chaos boundary layer structure of the FT
encoding for switches between low lying states,
which parallels the one found to encode for
rare thermal fluctuations \cite{chauve_creep,balentspld}.

We first investigate how two identical copies of a
disordered elastic system decorrelate when they
experience slightly different disorder at $T=0$.
Consider the hamiltonian:
\begin{equation}
H = \frac{1}{T} \sum_{i} \int d ^d x [ \frac{1}{2}  (\nabla u^i)^2 + V_i(u^i(x) ,x) ]
\label{ham}
\end{equation}
where the two copies $i=1,2$ are not mutually interacting. They
feel two different but mutually correlated random potentials
taken, with no loss of generality to
be gaussian with a correlation matrix:
\begin{eqnarray}
&& \overline{V_i(u ,x) V_j(u' ,x')} = R_{ij}(u-u') \delta^d(x-x')
\end{eqnarray}
and $R_{ij}(u) = R_{ij}(-u)$. We denote the
forces $F_i(x,u)=\partial_u V_i(u ,x)$. This is thus a two layer
version of the standard pinned elastic system, and the only
''coupling'' is due to the statistical correlations. For a 
short range (SR) function $R(u)$ it models an interface with random bond (RB)
disorder, for a long range (LR) function $R(u) \sim \sigma |u|$ at large u
it describes an interface in a random field (RF) of local variance
$\sigma$. For a periodic $R(u)$ it models a pinned one 
component CDW or Bragg glass, with generalizations to $N$-component
displacements $u_\alpha$.
We start with the case where all $R_{ij}(u)$ behave 
similarly at large $u$, e.g. $V_i=V \pm W \delta$, $V$ and $W$
being uncorrelated and identically distributed. The correlations are 
characterized by the matrix ($\langle .. \rangle$ are thermal averages):
\begin{eqnarray}
&& C_{ij}(x-x') = \overline{< (u^i(x) - u^i(x')) (u^j(x) - u^j(x')) >}  \nonumber
\end{eqnarray} 
The behaviour usually assumed from scaling arguments:
\begin{eqnarray}
&& C_{12}(x) = x^{2 \zeta} f(\delta x^\alpha) \label{scaling}
\end{eqnarray} 
corresponds to an overlap length $L_\delta \sim \delta^{-1/\alpha}$.
Here we show how to obtain (\ref{scaling}) from FRG
and compute the decay at large distance:
\begin{eqnarray}
&& C_{12}(x) \sim_{x \to \infty} x^{2 \zeta - \mu}
\label{decay}
\end{eqnarray}

We study the statics using the replicated Hamiltonian:
\begin{equation}
H_{rep} = \frac{1}{2 T}  \int d^dx \sum_{a i} (\nabla u^i_a)^2
- \frac{1}{2 T^2} \sum_{a b i j} R_{ij}(u^i_a - u_b^j) \label{hrep}
\end{equation}
One performs either Wilson RG 
varying the short scale momentum cutoff
$\Lambda_l=\Lambda e^{-l}$, or, equivalently to one
loop, compute the effective action of a uniform mode
adding a small mass (i.e. a term
$\frac{1}{2} m^2 u_i^2$ in (\ref{ham})) with $m_l=m e^{-l}$. Defining the
rescaled dimensionless correlator through
$R_{ij}(u) = A_d^{-1} m_l^{\epsilon - 4 \zeta} \tilde{R}_{ij}(u m_l^{\zeta})$,
with $A_4^{-1}=8 \pi^2$,
one derives the $T=0$ one loop FRG equations:
\begin{eqnarray} 
&& \partial_l \tilde{R}_{ij}(u) = (\epsilon - 4 \zeta) \tilde R_{ij}(u) + \zeta u  \tilde R'_{ij}(u)
\nonumber \\
&& + \frac{1}{2} \tilde R''_{ij}(u)^2 - \frac{1}{2} \tilde R''_{ij}(u) (\tilde R''_{ii}(0) +
\tilde R''_{jj}(0)) \label{1loop}
\end{eqnarray}
We also use their dynamical version, in terms of random force correlators 
$\tilde \Delta_{ij} = - \tilde R''_{ij}$, which, to one loop, describes
also the depinning threshold $f=f_c$, via
standard dynamical techniques \cite{chauve_creep,twoloopdep}.
Considering identical disorder in each layer,
$\tilde \Delta(u) \equiv \tilde \Delta_{12}(u)$ satisfies:
\begin{eqnarray}
&& \partial_l \tilde \Delta(u) = (\epsilon - 2 \zeta) \tilde \Delta(u)
+ \zeta u \tilde \Delta'(u) \nonumber
\\
&& - \tilde \Delta'(u)^2 - \tilde \Delta''(u)  (\tilde \Delta(u)  - \tilde \Delta(0))
+ \hat T \tilde \Delta''(u)
\label{frg1}
\end{eqnarray}
and we have denoted the positive quantity
\begin{eqnarray}
&& \hat T = \tilde \Delta_{11}(0) - \tilde \Delta(0) 
= \frac{1}{2} \overline{(F_1(0,x) - F_2(0,x))^2}
\label{temp}
\end{eqnarray}
measuring the difference between the
disorders in the two layers (at the bare level
$\hat T_{l=0} \sim \delta^2$). $\tilde \Delta_{11}(u)
=\tilde \Delta_{22}(u)$ obey identical uncoupled equations,
same as (\ref{frg1}) setting $\hat T = 0$, and
is known to develop a cusp and become non analytic 
beyond the Larkin length $R_c=e^{l_c}/\Lambda$.
$\hat T$ in (\ref{temp}) {\it formally} plays in the FRG equation
a role reminiscent to a ''temperature'' in rounding the
cusp at $u=0$. Its flow under RG
and its physics are however very different from temperature 
and must be here {\it determined self consistently} from (\ref{frg1},\ref{temp}). 
It is rather a static random force, which 
encodes static displacements below some scale,
and broadens the interlayer correlator preventing it from becoming
non analytic: $\Delta_{12}(u^t_1 + f_1 - u^{t'}_2 - f_2) \to
\frac{1}{2} \overline{(f_1 - f_2)^2} \Delta_{12}''(u^t_1 - u^{t'}_2 )$
with $f_i \equiv F_i(0,q)/q^2$ (in Fourier). This term arises because 
the statistical translational invariance (of
disorder under $u^i_a(x) \to u^i_a(x) + f_i(x)$)
present in each layer \cite{hwafisher},
does not hold for mutual displacements. There is no mutual pinning,
the two layers are
pinned independently, but there is some
residual correlation \cite{shocks}.

Let us analyze the flow equation (\ref{frg1})
for very small $\delta^2 \sim \hat T_0$. Below $R_c$ the non linear terms are
unimportant: the unrescaled $\Delta(0)$ and $\Delta_{11}(0)$
do not flow, $\hat T_l = e^{(\epsilon-2 \zeta) l} \hat T_0$
and $\tilde{\Delta}(u)$ and
$\tilde{\Delta}_{11}(u)$ remain approximately equal. Very
near $R_c$,
when $\hat T_l \tilde{\Delta}_{11}''(0) \sim \epsilon \tilde{\Delta}_{11}(0)$,
then $\tilde{\Delta}(u)$ starts to differ from 
$\tilde{\Delta}_{11}(u)$ within a 
boundary layer (BL) region of width $u \sim \hat T_{l_c} \ll 1$ around $u=0$ 
and remains analytic. Outside of this BL region, for
$u \sim O(1)$, $\tilde{\Delta}(u)$ is still almost equal to $\tilde{\Delta}_{11}(u)$
which then converges to the non-analytic FP function
$\tilde{\Delta}_{11}^*(u)$. The flow of $\tilde{\Delta}(u)$
beyond $R_c$, and its BL form, is much more difficult to study. 
For the case at hand (all $\tilde R_{ij}$ with similar
large $u$ behaviours),
it is clear that $\hat T_{l}$ keeps growing slowly.
Two important questions are thus (i) how to characterize this
growth, i.e. the overlap length 
(ii) where does it lead to, i.e. whether there is a finite 
$\hat T$ fixed point for $\tilde \Delta(u)$, i.e. 
are there residual correlations. We first address
(ii) and start with the periodic problem.

{\it 1- Periodic problem (CDW, Bragg glass)}: 
one assumes that fixed points
$\tilde \Delta_{11,l}(u)= \tilde \Delta^*_{11}(u)=\tilde \Delta^*_{22}(u)= 
\epsilon (\frac{1}{36} - \frac{1}{6} u (1-u)$
are reached within each layer. Using \cite{chauve_creep},
one finds that 
there is no non-trivial $\hat T >0$ fixed point.
Thus, $\tilde \Delta(u) = \tilde \Delta_{12}(u)$ flows 
slowly to zero as:
\begin{equation}
\partial_l \tilde \Delta(u) = \epsilon \tilde \Delta(u) 
+ (\epsilon/36) \tilde \Delta''(u)
\end{equation}
The leading decay, using potentiality $\int_0^1 \tilde \Delta(u) =0$, is
$\Delta_l(u) = g_l \cos(2 \pi u)$ with
$\partial_l g_l = \epsilon (1 -  \frac{(2 \pi)^2}{36} ) g_l$.
The correlation (\ref{scaling}) results from
the FRG flow of $\tilde \Delta$ from repulsive to attractive FP
and the general formula $C_{ij}(q)= \Delta_{ij,l}(0)/(q^2+m^2)^2$, exact for
$q=0$, $m=m_l$ and to $O(\epsilon)$ (one loop) accuracy 
for $q \sim m$ ($q=\Lambda_l$, $m=0$ for Wilson RG).
At large scale ($\zeta=0$ for the periodic problem):
\begin{equation}
C_{12}(q) \approx \tilde{\Delta}_{l,12}(0) q^{-4} e^{(2 \zeta
- \epsilon) l} |_{l=\ln(\frac{1}{q})} \sim
q^{- d - 2 \zeta + \mu} 
\label{general}
\end{equation}
Thus the
correlations of the displacements in the
two samples decay as in Eq. (\ref{decay}) with:
\begin{equation}
\mu = \epsilon ( (4 \pi^2/36) - 1) + O(\epsilon^2) \label{bg}
\end{equation}
For general periodic problem, e.g. the higher
component Bragg glass, to $O(\epsilon)$ one finds
$\mu = \epsilon (\tilde \Delta^*_{\alpha \beta}(0) K^0_\alpha K^0_\beta -1 )$
thus $\mu = \eta_{K_0} - \epsilon$,
where $\eta_{K_0}$ is the Bragg glass exponent
for decay of translational
order in a single copy $\overline{<e^{i K_0 u^1_x} e^{-i K_0 u^1_{x'}}>} \sim
x^{-\eta_{K_0}}$. One can compare the prediction (\ref{bg}) with the
one from Ref. \cite{hwafisher} in $d=2$ near $T=T_g$.
There, disorder generates an additional term
$\delta H_{rep} = \frac{-1}{2 T} \int_x [\sigma 
\nabla u^i_a \nabla u^i_b + 2 \hat \sigma \nabla u^1_a \nabla u^2_b]$ 
in (\ref{hrep}), with $\sigma_l$ growing as $B l$ and $\hat \sigma_l \to \hat \sigma^*$.
This implies that $C_{ii}(x) \sim B \ln^2 |x|$ while
$C_{12}(x) \sim \hat \sigma^* \ln |x|$, i.e. no absolute
decay of interlayer correlations ($\mu=0$), although decay relative to
intralayer ones. A numerical
study \cite{conum} is in progress to determine if $\hat \sigma^*>0$
at $T=0$ \cite{crossing}.

{\it 2- Short range (random bond) disorder}: 
There also we find that $\tilde \Delta(u)$ flows to
small values. Thus we study the linearized
part of (\ref{frg1},\ref{temp}) around $\tilde \Delta(u)=0$.
It is equivalent, and simpler, to study directly the
unrescaled linearized RG equation for $R=R_{12}$,
$\partial_l R(u) = - e^{\epsilon l} R_{11}''(0) R''(u)$. 
Since each layer reaches a fixed point, i.e.
$\tilde{R}_{11}(u) \to \tilde{R}^*_{11}(u)$, one has
$R_{11}''(0) \sim e^{(- \epsilon + 2 \zeta) l}  \tilde{R}_{11}^{* \prime \prime}(0)$. 
Thus $R_l(u)$ converges to a diffusion front ($R_{l=0}(u)$ being
peaked):
\begin{eqnarray}
&& R_l(u) \sim \frac{1}{\sqrt{4 \pi D_l}} e^{ - u^2/(4 D_l) } 
\int_{-\infty}^{+\infty} du' R_l(u') \label{rl}
\end{eqnarray}
with $D_l = \int^l_0 dl' e^{2 \zeta l'} \sim e^{2 \zeta l}$. 
Thus $\Delta_l(0)=- R_l''(0) \sim D_l^{- 3/2} \sim e^{- 3 \zeta l}$ and
from (\ref{general}) one finds (\ref{decay}) with:
\begin{eqnarray}
\mu=\mu_{RB} = 5 \zeta - \epsilon \label{musr}
\end{eqnarray}
always strictly positive (to one loop 
$\zeta_{RB}=0.28\epsilon$), i.e. the FP 
$\tilde \Delta=0$ is attractive.
This is easily generalized to the $O(N)$ model with the
result $\mu_{SR} = (N+4) \zeta - \epsilon$, replacing $R_{11}''(0)$ by $\partial_\alpha
\partial_\beta R_{11}(0) \sim \delta_{\alpha \beta}$ and considering diffusion in a
$N$-dimensional space. Finally, one shows that 
connected correlations of the free energies $F_i$
of each layer are given by 
$\overline{F_i F_j}^c=R_{ij}(0) L^d$ \cite{single}.
From (\ref{rl}) we find that
the decay of energy correlations is also of the form
$\overline{F_1 F_2} \sim L^{2 \theta-\mu_{RB}}$.

It is noteworthy that the replica variational method
\cite{mezpar} yields Flory exponents, i.e. $\mu=0$. 
The FRG captures the physics of the
decorrelations beyond the Flory approximation.
Since (\ref{musr}) results from linear FRG at $\tilde \Delta^*=0$,
one may look for simple interpretation.
Since each ground state satisfies
$\nabla^2 u^i(x) = F_i(u^i(x),x)$, residual
correlations in position can be interpreted
from correlation in pinning force resulting
from accidental encounters $u^1=u^2$ (i.e. averaging 
$\nabla^2 u^1 \nabla^2 u^2$, neglecting 
{\it all} correlations on the r.h.s. and using 
$u^i \sim L^\zeta$ yields (\ref{decay},\ref{musr})).
This argument (see also \cite{silveira}), is 
delicate however, since 
dimensional estimate (Flory) fail here, as
well as attempts to make it self-consistent.
In fact, linear analysis still involves non
trivial $\tilde \Delta_{11}^*$ fixed point
and (\ref{musr}) is confirmed by FRG
\cite{higherloop}. 

{\it 3- Long range (random field) disorder}: 
A different scenario occurs there. Using now the 
linearized unrescaled $\Delta(u)$ equation, i.e.
$\partial_l \Delta(u) = e^{\epsilon l} \Delta_{11}(0) \Delta''(u)$,
yields a diffusion form for $\Delta_l(u)$
with now $D_l \sim e^{2 \zeta l}$, thus $\mu = 3 \zeta - \epsilon$.
Since for the RF case $\zeta=\epsilon/3$ to all
orders (i.e. Flory is exact), the flow 
near $\tilde{\Delta}=0$ is {\it marginal} along
{\it a line of FRG fixed points}. These new FP describe the
residual (i.e. large scale) mutual correlations,
which we now describe. In the RF case, remarkably, 
the fixed point equation can be
integrated for any fixed $\hat T$ \cite{chauve_creep}.
Also the values of $\int_{0}^{+ \infty} du \Delta_{ij}(u) = - R'_{ij}(u=+\infty) 
= \sigma_{ij}$ are invariants (exactly to all orders) of the FRG flow,
where we denote $\sigma_{11}=\sigma$ and $\sigma_{12}=\sigma'=\sigma(1-\delta^2)$
the bare variances of inter and intra layer random fields. 
Defining rescaled variances $A_d \sigma= \epsilon \tilde \sigma$,
$A_d \sigma'= \epsilon \tilde \sigma'$, 
the function $\gamma(\tau) = \int_0^1 dy \sqrt{y - 1 - (1 + \tau) \ln y}$,
a reduced temperature $\tau=3 \hat T /(\epsilon \xi^2)$
and the two lengths $\xi_0^3 = \frac{3 \tilde \sigma}{\sqrt{2} \gamma(0)}$
and $\xi^3 = \frac{3 \tilde \sigma'}{\sqrt{2} \gamma(\tau)}$,
the fixed point functions are then:
\begin{eqnarray}
&& \tilde \Delta^*_{11}(u) = \frac{\epsilon}{3} \xi_0^2 y_0(u/\xi_0) \quad ,
\quad \tilde \Delta^*(u) = \frac{\epsilon}{3} \xi^2 y_\tau(u/\xi) \nonumber \\
&& y_\tau(x) - 1 - (1 + \tau) \ln y_\tau(x) = x^2/2
\end{eqnarray}
and $y_\tau(0)=1$. While $\tilde \Delta^*_{11}(u)$ is the standard
RF T=0 FP with a cusp non-analyticity at $u=0$,
$\tilde \Delta^*(u)$ has the
characteristic shape of a finite temperature FP
rounded by a (Chaos) BL of width $u \sim \hat T$.
To find this FP one needs to determine $\hat T$ (i.e. $\tau$)
for given RF variances $\sigma,\sigma'$.
It is given by the self consistency condition $(\ref{temp})$,
or equivalently $\xi^2 (1+ \tau) = \xi_0^2$. This yields:
\begin{equation}
 \sigma'/\sigma =  (1 + \tau)^{-3/2} \gamma(\tau)/\gamma(0)
\label{sc}
\end{equation}
which determines $\tau$ given $\sigma'/\sigma$, from which we obtain $\xi$,
then $\hat T$ and $\tilde \Delta^*(u)$. One sees from
(\ref{sc}) that there is indeed a fixed point $\tilde \Delta^*(u)$ for each value of
$\sigma'/\sigma$. In a sense
the overlap length is infinite, since there is a finite residual 
correlation (although there is a flow from infinitesimal $\hat T$
towards its FP value). It is described by universal functions,
e.g.:
\begin{eqnarray}
&& C_{12}(x)/C_{11}(x) \to_{x \to \infty} F_d[\sigma'/\sigma]
\label{resid}
\end{eqnarray}
with $F_d = (1+ \tau)^{-1/2}$ near $d=4$, $\tau$ being
solution of (\ref{sc}). It exhibits,
to one loop accuracy, a small $\delta$
singularity, 
from (\ref{sc}), $\tau \sim \delta/\sqrt{\ln(1/ \delta)}$.
An interesting quantity to compute numerically
is the ratio of center of mass fluctuations
(adding a mass $m$, i.e. harmonic well)
$C_{12}(q=0)/C_{11}(q=0) \to_{m \to 0}  G_d[\sigma'/\sigma]$
with $G_d=F_d$ near $d=4$ \cite{LR}. 

To discuss smaller scales ($\sim 
L_\delta$) we recall the droplet picture. Let
$u_1$ be the ground state (GS) of $V_1$.
A small SR perturbation $V_2-V_1 \sim \delta$ 
yields a new GS $u_2$. For small system size $L \ll L_\delta$
$u_1(x) \approx u_2(x)$ and the change in GS disorder energy 
is $\sim \overline{(E_1 - E_2)^2}^{1/2} \sim
\delta L^{d/2}$, where $E_i=\int d^d x V(x,u^i(x))$.
The (small) probability $p$ that another low energy state 
becomes more favorable inducing a (rare) large
switch $u_2-u_1 \sim L^\zeta$ is thus 
$p \sim \epsilon/L^\theta \sim \delta L^{(d/2) - \theta} \sim \delta
L^{\frac{\epsilon - 4 \zeta}{2}}$, as in thermal droplets.
$p$ becomes of order one at scale $L_\delta$. How does the FRG capture 
this physics ? Consider the growth at
short scale of $\overline{(F_1 - F_2)^2} \sim \Delta R L^{2 \theta}$
with $\Delta R= \tilde R_{11}(0) - \tilde R_{12}(0)$. 
From (\ref{1loop}) one finds $\partial_l \Delta R = (d - 2 \theta) 
\Delta R - \frac{1}{2} \hat T^2$. This suggests
that $\Delta R$ grows as $\delta^2 L^{d - 2 \theta}$ 
unaffected by the non linear term, until it 
reaches a FP value $O(1)$, and
reproduces the droplet result
at least for energy correlations.

However, more complex things could happen,
as seen on displacement correlations.
The GS difference $\overline{(u_1 -u_2)^2} \sim
\hat T L^{2 \zeta}$ is determined by the 
initial growth of $\hat T \sim L^a$. Within
droplets $\hat T$ is proportional
to $p$ the probability of rare GS switch,
and $a=a_{drop}=d/2 - \theta$. Rare GS switches induced
by thermal fluctuations was shown to be encoded in
the Thermal BL \cite{balentspld}. Here, we expect that the
Chaos BL (CBL) for $\tilde \Delta(u)$
describes switches induced by
disorder perturbations (higher moments of responses
being encoded in derivatives $\tilde \Delta^{(2n)}(0)$).
We found evidence for CBL solutions of the system 
(\ref{frg1},\ref{temp}) of the form 
$\tilde \Delta(u) = \tilde \Delta(0) + \hat{T} \phi(u/\hat{T}) + O(\hat{T}^2)$
and indications of a non trivial $a < a_{drop}$.
However, firm analytical as well as numerical 
determination of $a$ is {\it very} delicate \cite{duemmer}. A non trivial $a\neq a_{drop}$ 
arising from the CBL would require 
better understanding of minimization on all
scales. 

That (\ref{frg1},\ref{temp}) yields
a rich variety of behaviours is clear 
from e.g. the case of ''subdominant chaos''
$V_i(u) = V(u) \pm v(u)$ where $V(u)$ is
e.g. RF disorder and $v(u)$ RB disorder.
Then the ground state should not change at large scale 
since $\delta L^{d/2} \ll
L^{\theta_{RF}}$ with $\theta_{RF}=(d+2)/3$. 
$\hat T$ should spontaneously flow to zero (correlations
increase with scale) with $a_{drop}=d/2 - \theta_{RF}=
-\epsilon/6$ and a (self-organized) CBL solution
should form, similar to the TBL. Demonstrating
such RG flow, and testing universal
relations with thermal moments in $d=0$
is in progress.

Temperature chaos, i.e. two copies with different 
temperatures $T_i$, is described by 
Eq. (\ref{1loop}) adding the term $\frac{1}{2} (\tilde T_i + \tilde T_j) 
\tilde R''_{ij}$, with $\tilde T_i \sim T_i e^{-\theta l}$. 
The mutual correlator $\tilde \Delta$ still satisfies (\ref{frg1}) with:
$\hat{T} \to  \frac{1}{2} (\tilde T_1 + \tilde T_2) + \hat{T}$
where now $\hat{T}= \frac{1}{2} (\tilde \Delta_{11}(0) 
+ \tilde \Delta_{22}(0)) - \tilde \Delta_{12}(0)$.
Even for identical bare disorder, one finds that
although $\partial_l \hat{T}|_{l=0}=0$ one has
$\partial^2_l \hat{T}|_{l=0}= - \frac{1}{4} (T_1-T_2)^2 \Delta_{11}''''(0)$
thus an effective non zero $\hat{T}_{bare}$, and 
an independent {\it short range} $\tilde \Delta_{11}(u)-\tilde \Delta_{12}(u)$
is generated at small scales from entropic difference 
between the two copies. Apart from there being an additional scale
where the decreasing $\tilde T_i$ cross the
growing $\hat T$, the flow of $\tilde \Delta_{12}$
at larger scales, and around $L_\delta$
is similar (in the SR case) to disorder chaos. 

For dynamic chaos at depinning our result yields \cite{higherloop}
$\mu_{dep}  = 3 \zeta_{dep} - \epsilon$, with $\mu_{dep} >0$ 
from two loop corrections \cite{twoloopdep}, and thus slow decay of mutual correlations 
of critical configurations at $f_c$. 
However, it hold strictly in the fixed (or bounded) average center of mass 
ensemble, and a better treatment of the mode $q=0$ 
may be necessary \cite{depinning,periodicdep}.
Eq. (\ref{frg1},\ref{temp}) without the term $\tilde \Delta'(u)^2$
and $\epsilon=\zeta=0$, describes the {\it moving Bragg glass} in $d=3$.
One finds \cite{me} residual correlations as in (\ref{resid}), 
with $F_{d=3} \approx \int_0^1 du \Delta_{12}(u)^2/(2 \Delta_{11}(0)^2) + O(\delta^3)$
in terms of the bare disorder.

To conclude we applied FRG to 
chaos in pinned systems and obtained
the residual correlations.
In long range disorder these do not
decay and are described by new FRG fixed points.
The absence of chaos there
is reminiscent of weaker or absent chaos in 
infinite range models of more complex glasses.
Depinning and some quantum systems can be
studied by this method. The self-organized
chaos BL in the field theory uncovered here remains 
to be fully understood. We hope to stimulate 
numerical and experimental studies of 
correlations \cite{lrelasticity}. 

We thank K. Wiese, L. Balents, O. Duemmer, T. Giamarchi,
W. Krauth, C. Monthus, A. Rosso and G. Schehr for discussions 
and ongoing collaborations. 

\vspace{0.15cm}


\vspace{-0.2in}



\begin{thebibliography}{999}
\bibitem{crossing}
Mixing of $\nabla u^1_a \nabla u^2_b$ operator
(yielding a naive $\mu=d-2$) and FRG should
affect little accuracy of (\ref{bg}) in $d=3$.
\bibitem{higherloop}
Analysis of higher loops, though delicate,
suggests that (\ref{musr}) holds to all orders
with the exact value for $\zeta$.
\bibitem{LR} General LR disorder similarly
yields lines of FP. Another solvable case
is $\Delta_{ij}(u) \sim - \sigma_{ij} \ln|u|$,
$F_d=1-\sqrt{1 - \frac{\sigma_{12}}{\sigma_{11}}}$.
\bibitem{depinning}
some rare small change in disorder yield large forward jump 
and exploration of new environment. Chaos effects 
may thus be entangled with the statistics of avalanches.
\bibitem{periodicdep} 
In periodic depinning $\tilde \Delta_{11}(0) e^{\epsilon l}$ at the FP.
Residual correlations vanish unless $\int \tilde \Delta$ is corrected
in higher loop.
\bibitem{lrelasticity} 
Our result also holds for contact line elasticity.
\end{thebibliography}
\end{document}